\documentclass[aps,pre,twocolumn,showpacs]{revtex4}
\usepackage{epsfig}
\usepackage{times}
\usepackage{amsmath}

\begin{document}

\title{Defector-accelerated cooperativeness and punishment in public goods games with mutations}

\author{Dirk Helbing,$^{1,2,3}$ Attila Szolnoki,$^4$ Matja{\v z} Perc,$^5$ Gy{\"o}rgy Szab{\'o}$^4$}
\affiliation{$^1$ETH Zurich, CLU E1, Clausiusstr. 50, 8092 Zurich, Switzerland\\
$^2$Santa Fe Institute, 1399 Hyde Park Road, Santa Fe, NM 87501, USA\\
$^3$Collegium Budapest - Institute for Advanced Study, Szenth\'{a}roms\'{a}g u. 2, 1014 Budapest, Hungary\\
$^4$Research Institute for Technical Physics and Materials Science, P.O. Box 49, H-1525 Budapest, Hungary\\
$^5$Faculty of Natural Sciences and Mathematics, University of Maribor, Koro{\v s}ka cesta 160, SI-2000 Maribor, Slovenia}

\begin{abstract}
We study the evolution of cooperation in spatial public goods games with four competing strategies: cooperators, defectors, punishing cooperators, and punishing defectors. To explore the robustness of the cooperation-promoting effect of costly punishment, besides the usual strategy adoption dynamics we also apply strategy mutations. As expected, frequent mutations create kind of well-mixed conditions, which support the spreading of defectors. However, when the mutation rate is small, the final stationary state does not significantly differ from the state of the mutation-free model, independently of the values of the punishment fine and cost. Nevertheless, the mutation rate affects the relaxation dynamics. Rare mutations can largely accelerate the spreading of costly punishment. This is due to the fact that the presence of defectors breaks the balance of power between both cooperative strategies, which leads to a different kind of dynamics.
\end{abstract}

\maketitle

Evolutionary game theory formalizes the dynamics of populations of interacting individuals, considering the success (payoff) of their interactions. While this approach has applications in  biology \cite{sigmund_cb99,souza_mo_jtb09,moyano_jtb09,pacheco_prsb09,roca_plr09}, economics \cite{fehr_qje99}, and the social sciences \cite{santos_n08}, it has attracted a great deal of interest among physicists as well due to the relevance of methods from non-linear dynamics \cite{traulsen_09}, statistical physics \cite{szabo_pr07}, cellular automata \cite{nowak_n92b}, and many-particle simulations \cite{wu_zx_pre09}.
\par
One of the grand scientific challenges in this field concerns the question, how the outcome of interactions in social dilemma situations can be improved. In social dilemmas such as the public goods game, the collective well-being depends on the cooperation of individuals, which however is unlikely, as selfish behavior can generate higher personal profits. 
It has been proposed that reputation and costly punishment can fight free-riding (defection) and promote cooperation in public goods situations \cite{sigmund_pnas01}. It is puzzling, however, why people would make punishment efforts, as this reduces their payoffs compared to others who do not punish (``second-order free-riders''). In fact, punishing strategies disappear in public goods games, when the interactions between individuals are well mixed, creating again a tragedy of the commons. However, when individuals have spatial neighborhood interactions, free-riders may be eliminated (both, conventional and second-order ones)  \cite{helbing_ploscb10,brandt_prsb03}. This is due to the fact that the different cooperative strategies form clusters and segregate from each other. In this way, punishing cooperators avoid to be exploited by second-order free-riders (non-punishing cooperators) and can efficiently fight against defectors. Adding strategy mutations, however, endangers homogeneous clusters of individuals pursuing the same strategy, as they support the intrusion of competing strategies (``enemies''). For example, if defectors manage to enter a cooperative cluster, it can quickly erode. As a consequence, one would expect that mutations undermine the spreading of punishing strategies, thereby restoring the ``second-order free-rider problem'' and the ``tragedy of the commons''.
\par
Therefore, this paper investigates the impact of strategy mutations on the evolution of cooperation in the spatial public goods game with punishing strategies. As interaction graph, we assume a square lattice. Punishment is introduced by means of two additional strategies besides cooperators (C) and defectors (D). These two strategies are punishing cooperators (PC) and punishing defectors (PD), both of which impose a fine on defectors at a personal cost. 
The public goods game is iteratively played on a fully occupied square lattice of size $L \times L$ with periodic boundary conditions, where each player $x$ holds a strategy $s_x\in\{\mbox{C, D, PC, PD}\}$. Initially, the four strategies are equally and uniformly distributed over the $L^2$ lattice sites. Each player $x$ is a member of $G=5$ groups consisting of 5 individuals each. Each of these groups corresponds to a Neumann neighborhood of the focal individual or one of the direct neighbors.
\par
In each iteration, an individual $x$ plays a public goods game in all groups it belongs to. Cooperative individuals (playing C or PC) make a contribution of 1, while non-cooperative individuals (D or PD) contribute nothing. Afterwards, the sum of all contributions in each group is multiplied with the ``synergy factor'' $r$, and the resulting amount is equally shared between all of its members, irrespective of their contribution.
\par
Let $P_x^*$ denote the sum of the shares that individual $x$ receives in all of the $G$ groups it participates in. Then $P_x^*$ corresponds to the overall payoff of individual $x$ in the absence of punishment. This payoff is modified by punishment fines and punishment costs as follows: If $s_x =$ D or PD, player $x$ is punished with a fine $f$ in such a way that the remaining payoff is $P'_x=P_x^{*} - \sum f \pi_p$. Herein, the sum runs over all the groups containing player $x$. $\pi_p$ is given by the number of punishers (PC or PD) in each group (not considering player $x$), divided by $G-1$. Moreover, if $s_x =$ PC or PD, player $x$ invests a punishment cost $c$ such that the finally remaining payoff is $P_x=P'_x - \sum c \pi_d$. Herein, the sum runs again over all the groups containing player $x$. $\pi_d$ is given by the number of defecting individuals around player $x$ in each group (D or PD), divided by $G-1$. The division by $G-1$ serves to scale for the group size $G$.
\par
The strategies are updated according to the following Monte Carlo procedure: In each elementary step, a player $x$ and one of its neighbors $y$ is randomly chosen. For both individuals, the payoffs $P_x$ and $P_y$ are determined as described above. It is assumed that individual $y$ imitates the strategy $s_x$ of individual $x$ with probability $W = {\{1+\exp[(P_y-P_x)/K]\}}^{-1}$,
where $K$ denotes the uncertainty of strategy adoptions \cite{szabo_pre98}. 
Here, we use the value $K=0.5$. During one full iteration (Monte Carlo step MCS), the strategy of each player may be copied once on average.
\par
Following the work of Traulsen et. al. \cite{traulsen_pnas09}, mutation is introduced as a separate process. Accordingly, a player changes his or her strategy randomly (independently of the neighborhood) with a probability $\mu$, while the above described strategy adoption process is executed with probability $1-\mu$. In other words, in the limit $\mu \to 1$, the game-specific strategy adoption is completely ignored.
\par
Initially, each player follows a strategy at random. For all combinations of cost and fine parameters, the simulations were performed for systems with $L\ge 400$. Values greater than 400 (up to 1600) were chosen in the vicinity of the phase boundaries. This served to avoid that small strategy clusters would disappear by accident (by chance). The fractions $\rho_s$ of individuals using the strategies $s$ were determined after the transient time (up to $10^6$ iterations, depending on the speed of convergence).

\begin{figure}
\centerline{\epsfig{file=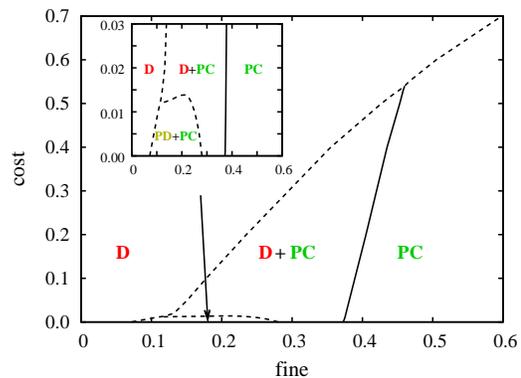,width=7.5cm}}
\caption{(color online) Phase diagram of the spatial public goods game with cooperators (C), defectors (D), and two punishing strategies, PC and PD, in the absence of mutations ($\mu=0$) for a synergy factor of $r=3.5$ (after \cite{helbing_ploscb10}). The inset magnifies the part of the phase diagram for small cost values, where the two punishing strategies PC and PD can coexist. Dashed lines indicate a first-order, solid lines a continuous phase transition.}
\label{phase}
\end{figure}

In the absence of punishment and mutation, cooperators die out at $r=3.74$, as can be concluded from Fig.~2 of Ref. \cite{szolnoki_pre09c}. For lower synergy factors, defectors dominate, while for higher values of $r$, cooperators can survive, or even spread all over the system (if $r>5.45$). Taking these values as a reference, Fig.~\ref{phase} shows a representative phase diagram of the spatial public goods game with punishment. As the punishment fine $f$ is increased, it can be observed that (for intermediate values of the punishment cost $c$) the system goes from a pure D phase over a mixed D+PC phase to a pure PC phase. If the cost of punishment is high ($c>0.51$), the mixed D+PC phase disappears completely, and the system directly changes from a pure D to a pure PC phase via a discontinuous phase transition. In the other extreme, if the cost of punishment is low ($c<0.013$), we have an additional area characterized by a coexistence of PC and PD (see inset of Fig.~\ref{phase}). Quite surprisingly, the second-order free-rider strategy C is not sustainable for $r=3.5$. Only if $r$ is increased, the pure D phase becomes a mixed D+C phase, which is the only phase where non-punishing cooperators can survive \cite{helbing_ploscb10}. For lower values of $r$, the mixed D+PC phase vanishes altogether, thus leaving the pure D and the pure PC phases as the only sustainable solutions, with a discontinuous transition between both phases when a critical $c(f)$ line is crossed \cite{helbing_ploscb10}.

The problem of second-order free riders results from the fact that pure cooperators bear no punishment cost, while receiving the same share of the public good as punishing cooperators (given the spatial strategy configuration is the same). This is, why non-punishing cooperators (``second-order free-riders'') crowd out punishing one under well-mixed conditions. However, the resulting tragedy of the commons is naturally resolved in structured populations \cite{helbing_ploscb10,brandt_prsb03}. There, the victory of the punishing cooperators is not based on a direct competition between the C and PC strategies, but rather on their different success in encounters with defectors. Due to the fixed, finite neighborhood, both the PC and C strategies form homogeneous clusters on the spatial grid and are exploited by defectors. If the fine is sufficiently large, punishing cooperators can overcome defectors, while cooperators can not. (Remember that $r>3.74$ is needed for cooperators to be sustainable in the presence of defectors.) Thus, punishing cooperators can spread when competing with defectors, while non-punishing cooperators are crowded out by them. As a consequence, second-order free-riders (cooperators) disappear, while punishing cooperators take over.

\begin{figure}
\centerline{\epsfig{file=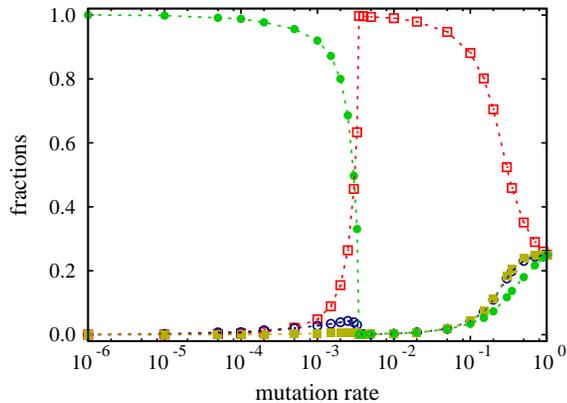,width=7.5cm}}
\caption{(color online) Fractions of all four strategies in dependence of the mutation rate $\mu$ for $c=0.6$ and $f=0.6$, which for $\mu=0$ lies in the PC phase. Filled green circles represent punishing cooperators (PC), open blue circles cooperators (C), filled yellow squares punishing defectors (PD), and open red squares defectors (D).}
\label{pcmutate}
\end{figure}

In the following, we investigate how robust this mechanism based on the clustering and segregation is with respect to strategy mutations. We proceed similarly as in Ref. \cite{traulsen_pnas09}, but for a spatial setting and considering punishing defectors rather than loners. For each phase displayed in Fig.~\ref{phase}, we find the following typical behavior: Small mutation rates do not significantly change the strategy distribution as compared to the mutation-less case. However, for $\mu \approx 10^{-2}$ or higher, the fraction of defectors increases quickly to values close to 1, as mutations generate kind of well-mixed conditions, then. Finally, in the limit $\mu\approx 1$, mutations dominate the dynamics, leading to an equidistribution of strategies (i.e. the fraction of defectors drops again). Figure~\ref{pcmutate} shows a typical example for the PC phase, where the dominance of punishing cooperators is sustained until approximately $\mu \approx 10^{-3}$.

Naturally, the value of the mutation rate, beyond which defectors can efficiently spread, is highly dependent on the $f/c$-ratio. Increasing the fine $f$ can reduce the impact of mutations, because this strengthens punishing cooperators compared to defectors. Nevertheless, sufficiently high value of $\mu$ eventually promote the spreading of defectors through the creation of a kind of well-mixed state. As emphasized before, a successful spatial clustering and segregation of strategies is a precondition for the spreading of cooperative behavior and punishment in the public goods game.

\begin{figure}
\centerline{\epsfig{file=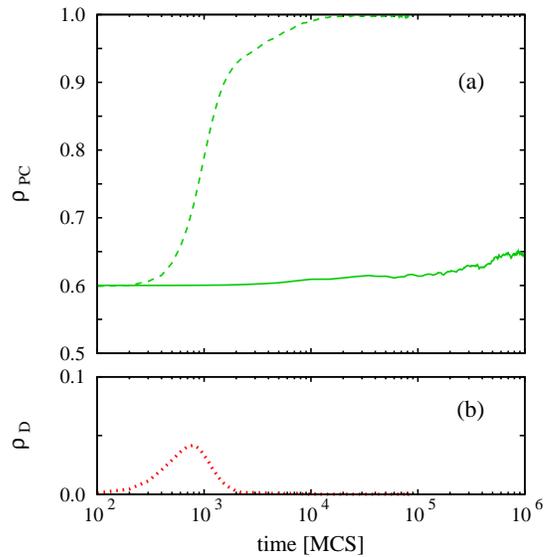,width=7.5cm}}
\caption{(color online) Time evolution of the fraction $\rho_{\rm PC}$ of punishing cooperators 
for a small mutation rate of $\mu=10^{-6}$ (dashed green line in panel (a)), and in the absence of defecting strategies and mutations (solid green line). In both cases, the initial state was assumed to consist of $60\%$ punishing cooperators and $40\%$ cooperators. Panel (b) shows the time evolution of the fraction $\rho_{\rm D}$ of defectors in the presence of small strategy mutations ($\mu=10^{-6}$). Curves are averages over $100$ independent runs for a grid of size $1600 \times 1600$.}
\label{voter}
\end{figure}

While the introduction of moderate mutations does not significantly affect the final {\it outcome} of the competition between strategies, this does not necessarily apply to the \textit{dynamics}, particularly when both cooperative strategies (PC and C) become equivalent after the extinction of defecting strategies (D and PD). When non-punishing compete with punishing cooperators, a slow logarithmic coarsening (in the absence of surface tension) takes place, which is equivalent to the dynamics of the voter model. Despite the slow dynamics, the fixation to the absorbing PC phase is relatively fast, because, after the extinction of defectors, the fraction of punishing cooperators is high compared to the fraction of cooperators. This is a direct consequence of the greater success of punishing cooperators in the competition with defectors during the early stages of strategy competition, when the punishment cost is large enough. However, if the fractions of punishing and non-punishing cooperators were about the same and no defecting strategies were present, it would require exceptionally long to reach any of the absorbing states (C-only or PC-only). Such a  scenario is illustrated in Fig.~\ref{voter} (solid green line), where the initial fraction of punishing cooperators is assumed to be 0.6 and the fraction of cooperators is assumed to be 0.4.

Remarkably, even if a tiny mutation rate is introduced, which occasionally creates defectors and punishing defectors, this generates an enormous advantage of punishing cooperators over cooperators in the battle with defecting strategies. The presence of defecting strategies destroys the equivalence of cooperators and punishing cooperators and breaks the balance of power in favor of punishing cooperators. This results in a striking acceleration of the spreading of punishing cooperators, as depicted by the dashed green line in Fig.~\ref{voter}. It is notable that a mutation rate as tiny as $\mu=0.000001$ evokes such an enormous change in the system dynamics.

\begin{figure}
\centerline{\epsfig{file=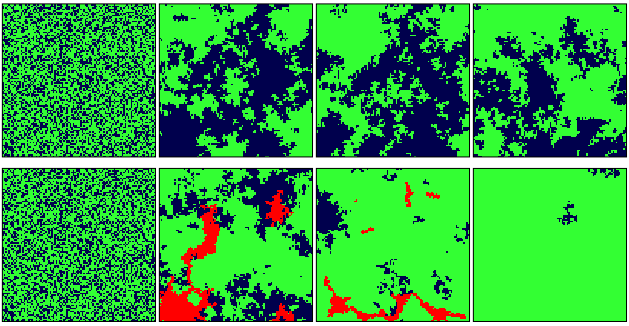,width=8.4cm}}
\caption{(color online) Typical snapshots of the simulation grid after 0, 900, 1,200, and 10,000 iterations during the coarsening process with strategy mutations (bottom) and without (top). The initial configurations are identical, and the parameter values agree with those of Fig.~\ref{voter}. All panels show a $100\times 100$ part of the $1600\times 1600$ grid. Red (grey) sites correspond to defectors, green (light) ones to punishing cooperators, and blue (dark) ones to cooperators. Punishing defectors cannot survive. It can be clearly seen that the presence of defectors due to strategy mutations largely accelerates the spreading of punishing cooperators.}
\label{snapshots}
\end{figure}

The mutation-induced acceleration of the coarsening process relies on the same effect that creates the dominance of punishing cooperators over cooperators. It gives rise to a D+PC phase and, solves the second-order free-rider problem due to the disappearance of non-punishing cooperators. When defectors occur in the vicinity of cooperators, they can spread efficiently because of the low value of $r$. Defectors, however, cannot succeed against punishing cooperators, if the fine is sufficiently high. Consequently, punishing cooperators spread at the cost of defectors, while these crowd out cooperators. As a consequence, a quick victory of punishing cooperators over cooperators requires an interaction between \textit{three} strategies: C, PC and D. The snapshots of Fig.~\ref{snapshots} demonstrate the coarsening process impressively for the cases with mutation (bottom row) and without (top row). Starting with identical spatial distributions of cooperators and punishing cooperators, it can be observed that the fractions of the two strategies remain almost the same, when no mutations take place. However, in the presence of a small rate of strategy mutations ($\mu = 10^{-6}$), defectors can temporarily spread in the population at the expense of cooperators. This, in turn, provides conditions for a fast spreading of punishing cooperators at the expense of defectors. This dynamics replaces the slow logarithmic coarsening in the absence of defectors. Notably, once punishing cooperators take over the majority of the spatial grid, a significant fraction of defectors can no longer exist. This can be seen on the bottom of Fig.~\ref{snapshots} as well as in panel (b) of Fig.~\ref{voter}, which demonstrates the temporary uprise of defectors just before punishing cooperators prevail in the system.

In summary, we have studied the evolution of cooperation in public goods games with mutation and punishment, where punishing cooperators and punishing defectors were taken into account besides conventional cooperators and defectors.  Considering structured populations naturally solves the second-order free-rider problem by spatially separating the interaction of cooperators with defectors and of defectors with punishing cooperators. Since punishing cooperators are able to outperform defectors at sufficiently large punishment fines and defectors are superior to cooperators, punishing cooperators are the winners of the strategy competition in space. This mechanism is robust to modest mutation rates, while large mutation rates create kind of well-mixed interactions, which promote a spreading of defectors and, thereby, a tragedy of the commons. Naturally, in the limit $\mu \to 1$ mutations become so strong that they create a game-independent random strategy distribution.

Despite the robustness of the final outcome to moderate mutation rates, we could demonstrate that even tiny mutation rates can have an enormous impact on the evolutionary dynamics, particularly when punishing cooperators would otherwise compete with cooperators only. When no other strategies are present, punishing and non-punishing cooperators receive the same payoffs, which leads to a slow logarithmic coarsening as in the voter model. The occurrence of defectors through strategy mutations breaks the balance of power between the two cooperative strategies. This can dramatically accelerate the spreading of punishing cooperators. Note that, in many systems, mutations lead to a different outcome. In the model studied here, however, mutations have an effect like a catalyst: they speed up a process while the outcome of the system is not affected.

We acknowledge partial financial support by the Future and Emerging Technologies programme FP7-COSI-ICT of the European Commission through the project QLectives (grant no.: 231200) and by the ETH Competence Center ``Coping with Crises in Complex Socio-Economic Systems'' (CCSS) through ETH Research Grant CH1-01 08-2 (D.H.), by the Hungarian National Research Fund (grant K-73449 to A.S. and G.S.), the Bolyai Research Grant (to A.S.), the Slovenian Research Agency (grant Z1-2032-2547 to M.P.), and the Slovene-Hungarian bilateral incentive (grant BI-HU/09-10-001 to A.S., M.P. and G.S.).

\end{document}